\documentclass[12pt,a4paper]{article}
\usepackage{amsmath, amssymb}
\usepackage{amsthm}
\usepackage{enumerate}
\usepackage{graphicx}
\usepackage{color}
\usepackage{comment}
\usepackage{geometry}
\usepackage{setspace}
\usepackage{url}
\usepackage{newtxtext,newtxmath}
\usepackage{titletoc}
\usepackage{tikz}
\usetikzlibrary{decorations.pathreplacing}

\usepackage[colorlinks = true, citecolor = blue, linkcolor = black]{hyperref}
\usepackage{natbib} 

\newtheorem{proposition}{Proposition}

\newfont{\bg}{cmr9 scaled\magstep4}
\newcommand{\bigzerol}{\smash{\lower1.0ex\hbox{\bg 0}}}

\begin{document}

\title{{\bf Downsian Competition for the Myerson Value}\footnote{I would like to thank Satoshi Kasamatsu, Satoshi Nakada, and Susumu Sato for their helpful comments. This study was financially supported by JSPS KAKENHI (Grant Number: 26K16320).}}
\author{Daiki Kishishita\thanks{Graduate School of Economics, Hitotsubashi University, Japan.  2-1 Naka, Kunitachi, Tokyo, Japan. 186-8603. E-mail: \url{daiki.kishishita@gmail.com}. } }

\date{\today}

\maketitle
\begin{abstract}
This paper studies an electoral competition model in which parties maximize legislative power rather than vote shares. Voters are uniformly distributed on the unit interval and vote for the party proposing the closest policy platform. After the election, parties form coalitions through a communication network arising from ideological proximity: two parties are directly linked if their policy distance is at most $d$. A party's objective is its Myerson value in the resulting graph-restricted voting game. I characterize symmetric pure-strategy equilibria in two-, three-, and four-party systems. The two-party case yields convergence to the median. The three-party case admits a continuum of symmetric equilibria in which the two extreme parties are directly linked. In the four-party case, the unique symmetric equilibrium places two parties at $(1-d)/2$ and two parties at $(1+d)/2$. In both three- and four-party systems, more restrictive coalition communication, represented by a smaller $d$, generates a centripetal force, and the median voter theorem holds in the limit despite the multiparty setting.
\begin{description}
    \item[Keywords:] Downsian competition; Myerson value; multiparty system; coalition; policy divergence
    \item[JEL classification codes:] D72; C70
\end{description}
\end{abstract}

\newpage
\begin{spacing}{1.2}

\section{Introduction}
The Downsian model treats parties as vote maximizers and predicts convergence toward the
median voter in the canonical two-party system \citep{downs1957economic}. In a multiparty parliamentary
system, however, electoral support is only one determinant of political influence. A party’s
platform also affects whether it is viewed as an acceptable coalition partner and, consequently,
how pivotal it is in government formation. Policy choice may therefore change both a party’s
seat share and its bargaining position after the election.

This second channel is empirically relevant. Studies find that greater policy distance tends to reduce the probability of coalition formation \citep[e.g.,][]{martin2001government,budge2016party}, and electorally significant parties at the ideological extremes are often excluded from government cooperation through a \textit{cordon sanitaire} \citep{axelsen2024cordon}. The Sweden Democrats, for example, were long excluded from direct government cooperation even after becoming important for parliamentary majorities \citep{backlund2023government}.\footnote{The analogous norm against cooperation with the AfD in Germany is commonly described as a ``firewall'' (\textit{Brandmauer}) \citep{kirby2025germanyfirewall}.} These observations raise a natural theoretical question: how does the pursuit of post-electoral coalition power alter parties' platform choices? Yet the implications of coalition-bargaining incentives for electoral competition remain underexplored.

I address this question by combining a standard spatial voting model with the Myerson value \citep{myerson1977graphs}. Parties choose positions on a one-dimensional policy space. Voters support the closest party, so positions determine proportional seat shares. The same positions also generate a communication graph: two parties are directly linked if and only if their policy distance is at most a threshold $d$. Direct and indirect links determine which groups of parties can act jointly, and each party's payoff is its Myerson value in the induced graph-restricted voting game. Policy choice therefore affects legislative power through two endogenous objects---vote share and coalition connectivity.

The Myerson-value objective has a transparent bargaining interpretation. The Shapley--Shubik index measures a party's expected marginal contribution to winning coalitions when the order in which parties enter negotiations is random \citep{shapley:book1952,shapley1954method}.\footnote{Since then, a variety of power indices have been proposed \citep{napel2018voting}.} The Myerson value applies the same logic after restricting cooperation by a communication graph. It is therefore a natural reduced-form objective for office-seeking parties that choose platforms to maximize expected bargaining leverage rather than votes per se. Importantly, the model does not impose coalition blocs exogenously: the communication structure is itself generated by policy choices.


To measure how much policy distance can be bridged in coalition negotiations, I introduce a parameter \(d \in(0, 1/2)\), which I call the coalition-communication threshold. I assume that two parties are directly linked if and only if their policy distance is at most \(d\).   On the one hand, a larger $d$ allows direct communication between more distant parties.  In such a case, parties that are far apart in policy space find it difficult to negotiate or form a coalition directly, while an ideologically intermediate party may  serve as a mediator between them \citep{napel2012monotonicity}. On the other hand, a smaller $d$ represents a more restrictive environment, as may arise when ideological conflict is salient, compromise is costly, or cross-party communication is difficult. One might expect a smaller $d$ to push parties farther apart. The central result is the opposite: tighter communication constraints can pull parties toward the center.

Assuming that parties maximize the Myerson value, this paper analyzes policy  platforms in electoral competition. I characterize symmetric pure-strategy equilibria in two-, three-, and four-party systems. The analysis shows that the effect of coalition-bargaining incentives depends sharply on the number of parties, and the coalition-communication
threshold, $d$, shapes electoral competition in the multiparty system.

First, in the two-party system, the unique symmetric equilibrium is full convergence to the median voter. Thus, incorporating coalition-bargaining incentives does not change the standard Downsian prediction in the two-party case. This is intuitive: when there are only two parties, coalition formation is not a substantive concern, and the usual logic remains.

The conclusion changes in multiparty systems. In these cases, the coalition-communication threshold \(d\) plays a central role. The threshold captures how easily policy proximity translates into coalition communication. When \(d\) is small, even relatively close parties may be unable to communicate or cooperate. This may correspond, for example, to a political environment in which policy issues are highly salient, ideological compromise is costly, or inter-party polarization makes communication across political lines difficult.

In the three-party system, a continuum of symmetric equilibria exists: any profile
\[
(s,1/2,1-s) \quad 
{\rm with} \quad
\frac{1-d}{2}\le s\le \frac12
\]
is an equilibrium. This contrasts with the standard Downsian model, in which no pure-strategy equilibrium exists with three vote-maximizing parties. In the present model, coalition communication creates an additional strategic consideration. As long as even the two extreme parties are close enough to communicate directly, no party can profitably change its Myerson value by unilateral deviation. As $d$ decreases, the set of equilibria shrinks toward full convergence at the median. In the
limit, as $d\to 0$, the median voter theorem holds despite the multiparty setting.

The four-party system delivers the main result. The unique symmetric equilibrium is
\[
\left(
\frac{1-d}{2},\frac{1-d}{2},
\frac{1+d}{2},\frac{1+d}{2}
\right).
\]
Thus, the distance between the two party pairs is  equal to the coalition-communication threshold. If \(d\to 1/2\), the equilibrium coincides with the usual four-party Downsian benchmark,
\[
\left(\frac14,\frac14,\frac34,\frac34\right).
\] By contrast, as \(d\to 0\), the equilibrium platforms converge to the median voter's ideal point. In this sense, the median voter theorem again holds in the limit despite the multiparty setting.

At first glance, a smaller coalition-communication threshold may seem likely to induce greater policy divergence because policy compromise becomes more difficult. The analysis shows the opposite. More restrictive coalition communication generates a centripetal force that is absent from the standard vote-maximization benchmark. When communication is difficult, parties have an incentive to move closer to the center in order to remain connected to other parties and preserve their bargaining power in government formation. 

As an application, we also examine the four-party system with two extreme parties. We find that a coalition between the two mainstream parties emerges endogenously in equilibrium because of their strategic
policy choices to maximize the Myerson value. This is consistent with the cordon sanitaire mentioned at the beginning. Furthermore, as $d\to 0$, the median voter theorem again holds for the mainstream parties. 

These results highlight the importance of explicitly considering coalition-power incentives when predicting policy platforms in multiparty parliamentary systems.

\paragraph{Related literature}
This study is related to several strands of literature. First, it contributes to the literature on power indices in voting games. Various studies have applied the Myerson value or its variants to voting games \citep[e.g.,][]{vazquez1996owen,calvo1997probabilistic,
napel2012monotonicity,skibski2017algorithm,koki2019coalitions}.\footnote{More broadly, the Myerson value is an index for graph-restricted voting games. Using a different type of power index for this class of games, \cite{benati2021voting} analyze bargaining within the Council of the European Union.} These studies take the distribution of seats and the communication structure as given and measure each party's power. In contrast, this paper endogenizes both vote shares and the communication graph through parties' policy choices. The analysis therefore shows how the pursuit of power measured by the Myerson value feeds back into electoral competition.

Second, the paper is related to the literature on multiparty electoral competition. Multiparty competition has been analyzed under various assumptions, including vote-maximizing parties \citep[e.g.,][]{merrill2002centrifugal, evrenk2009three}, broader vote-related objectives including vote-share maximization, plurality-margin maximization, and complete plurality maximization \citep{cox1987electoral}, parties caring only about the winning probability \citep{chisik2006winning},
and policy-motivated parties \citep[e.g.,][]{matakos2016electoral}.
 Studies using probabilistic voting models show policy convergence toward the mean, rather than the median,  in multiparty competition with vote-maximizing parties \citep[e.g.,][]{adams1999multiparty}. The present paper differs from these studies by assuming that parties maximize coalition-bargaining power rather than vote shares or policy payoffs. A novel implication is that convergence to the median can emerge in the limit under deterministic voting.

Third, our analysis is related to the literature on network formation. Our assumption that a link is formed if and only if the policy distance between two parties is no greater than $d$ is conceptually similar to that of \citet{iijima2017social}, who study network formation under the assumption that two agents form a link if and only if the distance between their characteristics is below a given threshold. \citet{morelli2004party} employs a similar network structure to study party formation rather than government formation. Neither study, however, uses the Myerson value to determine agents’ payoffs under the network structure. By contrast, \citet{aumann2003endogenous} analyze an endogenous network-formation model in which realized payoffs are determined by the Myerson value. A key difference from their framework is that, in our setting, policy choices affect not only the network structure but also parties’ vote shares, which jointly determine each party’s Myerson value.

Finally, the study by \cite{austen1988elections} is particularly relevant to this paper.\footnote{\cite{troumpounis2016incomplete} also analyze three-party electoral competition with government formation, but they do not allow government formation to depend directly on policy platforms. Instead, their focus is strategic voting by voters, which is ignored in this study.} They analyze how government formation affects electoral competition. The present study differs from theirs in several respects. First, I consider both the three-party and four-party cases, whereas they consider only the three-party case. Since a pure-strategy equilibrium does not exist in the standard Downsian framework with three parties, the effect of coalition formation on party positioning is difficult to isolate in their setting. Second, I adopt a cooperative-game approach, whereas they use a non-cooperative bargaining model. Non-cooperative bargaining models require a specification of the bargaining protocol. By contrast, the Myerson value summarizes parties' bargaining power directly from the communication graph. This allows me to isolate the effect of coalition feasibility on electoral competition and to study comparative statics with respect to the coalition-communication threshold $d$. Indeed, their model does not include a parameter corresponding to  \(d\). Therefore, it cannot generate the main comparative statics studied in this paper.

\section{Model}
\subsection{Electoral Competition}
I consider the standard Downsian model except for the parties' objectives.

Fix \(n\in\{2,3,4\}\), and let \(N=\{1,\ldots,n\}\) be the set of parties. Each party \(i\in N\) simultaneously chooses a policy position \(x_i\in[0,1]\) at the beginning of the game. 

Each voter has a single-peaked preference with an ideal point. Voters' ideal points are uniformly distributed on \([0,1]\), which is known by the parties. Thus, the median voter's ideal point is $1/2$. Each voter votes for the party whose policy is closest to the voter's ideal point.\footnote{Voters may cast their votes strategically in order to affect government formation \citep[e.g.,][]{austen1988elections,baron2001elections,troumpounis2016incomplete}. I have to abstract from this possibility because my reduced-form approach to measuring bargaining power does not specify which policy is implemented for a given distribution of vote shares.
} If several parties are tied for being nearest, the voter votes randomly among them. 

Let \(m_i(x)\) denote party \(i\)'s vote share at \(x=(x_1,\ldots,x_n)\). The electoral system is proportional representation; thus, the fraction of seats obtained by a party is equal to its vote share.

Each party's objective is to maximize its Myerson value rather than the vote share, which will be defined in the next subsection.

\subsection{Communication Network and the Myerson Value}
This subsection introduces the concept of the Myerson value in the context of the present game structure.

\paragraph{Coalition formation as cooperative game} The legislative stage is represented by a weighted voting game. For any coalition \(S\subseteq N\), let
\[
m(S;x):=\sum_{i\in S}m_i(x),
\]
which represents the number of seats held by coalition \(S\). Define
\[
q(m(S;x)):=
\begin{cases}
1 & \text{if } m(S;x)>1/2\\
1/2 & \text{if } m(S;x)=1/2\\
0 & \text{if } m(S;x)<1/2
\end{cases},
\]
which is the probability that coalition \(S\) successfully forms a government.\footnote{Instead, we could consider the setting where \[
q(m(S;x)):=
\begin{cases}
1 & \text{if } m(S;x)>1/2\\
0 & \text{if } m(S;x)\leq 1/2
\end{cases},
\]
but the results do not change.
} The underlying voting game is given by the value of each coalition $S$:
\[
v(S):=q(m(S;x)).
\]

\paragraph{Communication network and Myerson value} A communication network is an undirected graph on \(N\) denoted by \(g\), where \(g\subseteq\{\{i,j\}\mid i,j\in N,\ i\neq j\}\). A pair \(\{i,j\}\in g\) represents a communication link between parties \(i\) and \(j\). Communication is constrained by policy distance. Given \(d\in(0,1/2)\),\footnote{If $d\geq 1/2$, even the extreme party with position 0 can be directly linked to the party with the median position, which is unrealistic. I impose this restriction to focus on substantively relevant cases. This restriction is crucial only for the four-party system; the same results for the two- and three-party systems hold as long as $d\in(0, 1)$.} define \(g\) as
\[
\{i,j\}\in g \quad \Leftrightarrow \quad |x_i-x_j|\leq d.
\]
Thus, two parties are directly linked if and only if their policy distance is at most \(d\).

Importantly, communication within a coalition does not require all parties in the coalition to be within distance \(d\). It is enough that they are connected through a chain of links. For example, parties \(i\) and \(k\) can communicate through party \(j\) if \(|x_i-x_j|\leq d\) and \(|x_j-x_k|\leq d\), even when \(|x_i-x_k|>d\). This captures the idea that parties that are far apart in policy space may
find it difficult to negotiate or form a coalition directly, but an ideologically intermediate
party may serve as a mediator between them.

For any coalition \(S\subseteq N\), let \(g|_S:=\{\{i,j\}\in g\mid i,j\in S\}\) be the restricted network of \(g\) on \(S\). $S$ is connected in $g|_S$ if there is a path from $i$ to $j$ in $g|_S$ for $i, j\in S$. In
particular, if $S$ is connected in $g$ and $S\cup\{i\}$ is not connected in $g$ for any $i\notin S$, we say that $S$ is a component of $g$. Furthermore, we denote by \(S/g\) the partition of \(S\) with respect to the components of \(g|_S\). That is, each $C\in S/g$ is a component of $g|_S$. 

\begin{figure}
    \centering
    \includegraphics[width=0.5\linewidth]{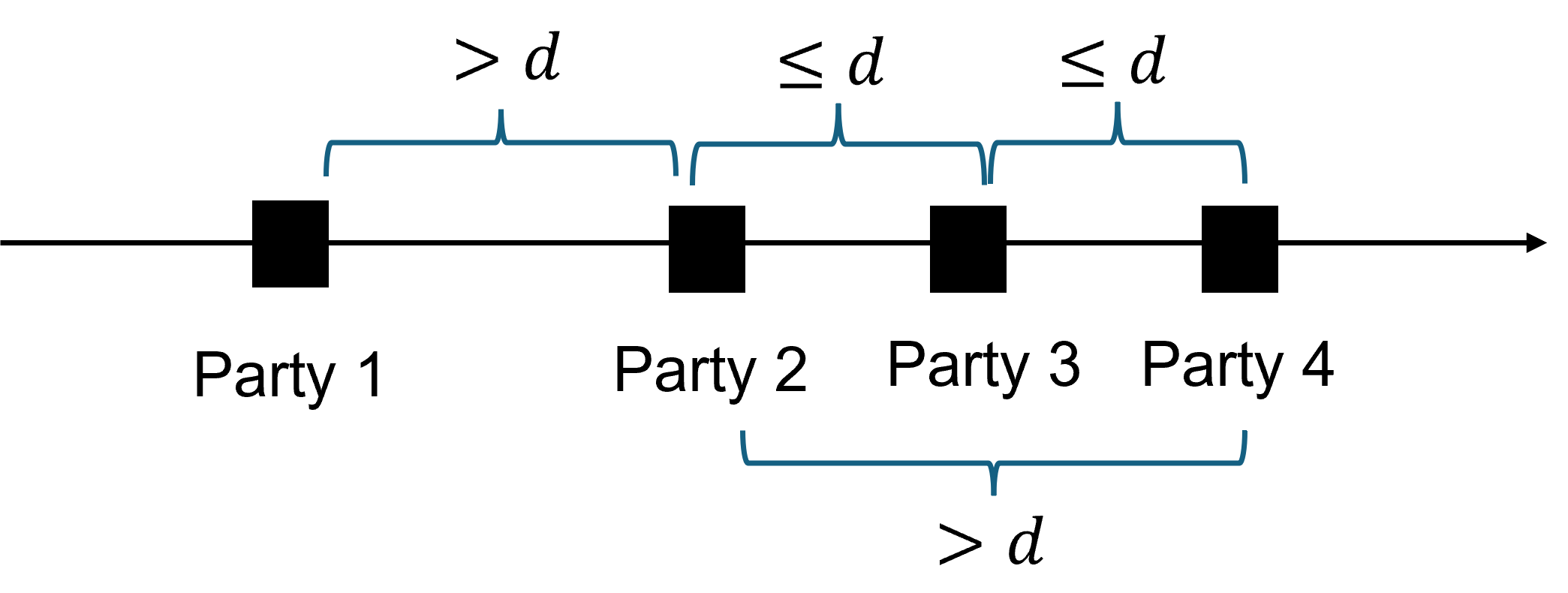}
    \caption{Example of restricted communication network}
    \label{fig:exam}
\end{figure}

To see these definitions, see Figure \ref{fig:exam} with four parties. In this case, $\{2, 3, 4\}$, $\{2, 3\}$, and $\{3, 4\}$ are connected, but for example, neither $\{1, 2, 3, 4\}$ nor $\{2, 4\}$ are connected. Thus, when $S=N$, $S/g=\{\{1\}, \{2, 3, 4\}\}$.

Then, the graph-restricted game is
\[
v^g(S):=\sum_{C\in S/g} v(C).
\]
That is, if a coalition \(S\) is connected, it can act as a single coalition. If \(S\) is disconnected, then only its connected components can cooperate internally, and the value of \(S\) is the sum of the values generated by those components.

Party \(i\)'s payoff is its Myerson value \citep{myerson1977graphs}, namely the Shapley value of the graph-restricted game:
\[
\mu_i
:=
\sum_{S\subseteq N\setminus\{i\}}
\frac{|S|!(n-|S|-1)!}{n!}
\left[
 v^g(S\cup\{i\})-v^g(S)
\right].
\]
\(v^g(S\cup\{i\})-v^g(S)\) captures the marginal contribution of party $i$ to coalition $S$. Thus, this value captures the average marginal contribution of party $i$ under the communication restrictions encoded by \(g\).

\paragraph{Example}
Finally, to illustrate how the Myerson value is computed, consider the four-party profile
\[
x=\left(\frac{1}{4},\frac{1}{4},\frac{3}{4},\frac{3}{4}\right)
\]
and suppose that \(d<1/2\). Then, parties 1 and 2 are linked, and parties 3 and 4 are linked, but there is no link between the two pairs. Thus, the communication graph has two connected components, \(\{1,2\}\) and \(\{3,4\}\). Since each party receives vote share \(1/4\),
\[
v^g(\{1,2\})=v^g(\{3,4\})=q(1/2)=\frac{1}{2},
\quad
v^g(\{1,2,3,4\})=v(\{1,2\})+v(\{3,4\})=1.
\]

Consider party 1. In the graph-restricted game, party 1 has a positive marginal contribution only when party 2 is already in the predecessor set. If party 2 is not in the predecessor set, adding party 1 creates only a singleton component with vote share \(1/4\), whose value is zero. If party 2 is in the predecessor set, adding party 1 creates the connected component \(\{1,2\}\), whose value is \(1/2\). Hence, party 1's marginal contribution is \(1/2\) for the predecessor sets
\[
\{2\},\quad \{2,3\},\quad \{2,4\},\quad \{2,3,4\},
\]
and zero otherwise. Therefore, party 1's Myerson value is
\[
\left(
\frac{1!2!}{4!}
+
\frac{2!1!}{4!}
+
\frac{2!1!}{4!}
+
\frac{3!0!}{4!}
\right)\frac{1}{2}
=
\frac{1}{4}.
\]
By symmetry, each party obtains the same Myerson value. 

Note that this does not imply that the sum of the Myerson values is always one. For example, suppose that all four parties are located at distinct positions and that no party has a majority on its own. As $d \to 0$, no party is connected to any other party. Since no single party can form a government, the value of every party is zero. This implies that electoral competition is not a zero-sum game, in contrast to the case of vote-maximizing parties. 

\subsection{Equilibrium Concept}
A pure-strategy Nash equilibrium is a policy profile \(x\) such that no party can increase its Myerson value by unilaterally changing its policy position. I focus on symmetric equilibria. In the two-party system, a symmetric profile is written as \((s,1-s)\), with \(0\leq s\leq 1/2\). In the three-party system, a symmetric profile is written as \((s,1/2,1-s)\), with \(0\leq s\leq 1/2\). In the four-party system, a symmetric profile is written as \((s,t,1-t,1-s)\), with \(0\leq s\leq t\leq 1/2\).

\section{Equilibrium}
I characterize the equilibria of this game sequentially for the two-, three-, and four-party systems.

\subsection{Two-Party System}
I start with the two-party system as a benchmark case. In this case, I obtain the following proposition. The proofs are contained in the Appendix.

\begin{proposition}\label{prp_two}
In the two-party system ($n=2$), the unique symmetric pure-strategy Nash equilibrium is
\[
x^*=\left(\frac12,\frac12\right).
\]
\end{proposition}
Therefore, the median voter theorem holds as in the Downsian framework. That is, coalition-building incentives do not alter the equilibrium prediction. Importantly, the coalition-communication threshold, $d$, has no impact on the equilibrium.

To see this result, consider the deviation incentive of a party from $1/2$ to a leftist policy. In the equilibrium, the Myerson value of each party is $1/2$. In contrast, the deviation makes the vote share of the party less than a half. Thus, the opponent always wins, meaning that the Myerson value of the party becomes zero. Therefore, as in the case of the standard Downsian model, there is no deviation incentive from the median voter's ideal point.

\subsection{Three-Party System}
However, this is no longer the case in the multiparty system. Regarding the three-party system, I obtain the following result.

\begin{proposition}\label{prp_three}
In the three-party system ($n=3$), the set of symmetric pure-strategy Nash equilibria is
\[X^*=
\left\{
\left(s,\frac12,1-s\right)
\ \middle|\
\frac{1-d}{2}\leq s\leq \frac12
\right\}.
\]
\end{proposition}

The proposition shows that the three-party case differs sharply from the two-party case. With two parties, both parties must converge to the median voter. With three parties, however, there is a continuum of symmetric equilibria. The communication constraint does not pin down a unique degree of policy divergence; rather, it determines the largest admissible distance between the two extreme parties.

The intuition is as follows. If \(1-2s\le d\), then the two extreme parties are directly linked, and hence the communication graph is complete. In this case, no party is individually winning, while every two-party coalition is winning. Therefore, the graph-restricted game is the standard three-player majority game, and each party obtains the same Myerson value, 1/3, regardless of the exact distribution of vote shares. A unilateral deviation cannot make a party individually majoritarian, and hence cannot yield a payoff greater than 1/3.

By contrast, if \(1-2s>d\), then the two extreme parties cannot communicate directly. In that case, at least one extreme party can profitably move inward and improve its communication position. Thus, symmetric equilibria require the two extreme parties to be close enough to communicate directly. 

As \(d\) decreases, the set of equilibria shrinks toward full convergence at the median. In the limit, as \(d\to 0\), the only symmetric equilibrium outcome converges to $(1/2,1/2,1/2)$. Thus, in the limit, the median voter theorem holds despite the multiparty setting.  

This result contrasts with the Downsian framework. When parties are vote maximizers, there is no pure-strategy equilibrium; in the mixed-strategy equilibrium, each party randomizes uniformly over $[1/4,3/4]$ \citep{shaked1982existence}. By contrast, when parties maximize the Myerson value, pure-strategy equilibria exist, and the median voter theorem holds in the limit.\footnote{This does not imply that realized policy divergence is always smaller when parties maximize the Myerson value than when they maximize vote shares. In the Downsian mixed-strategy equilibrium, parties may happen to choose very close platforms, which makes it difficult to compare realized policy divergence across the two models. This difficulty does not arise in the four-party system examined in the next subsection, because both models admit pure-strategy equilibria.
} Note that the result also contrasts with the case where parties care only about winning the election. In this case, the median voter theorem never holds, although pure-strategy equilibria exist \citep{chisik2006winning}.

\subsection{Four-Party System}
I now turn to the main case of interest: the four-party system. I regard this as the main interest for two reasons. First, four parties provide the minimal setting in which two ideologically proximate parties may form a government, while cross-bloc cooperation between a left-leaning and a right-leaning party is also possible. Second, unlike the three-party Downsian model, the four-party benchmark admits a pure-strategy equilibrium, which enables us to cleanly compare the equilibrium platforms between the two cases.

In the four-party system, I obtain the following proposition:

\begin{proposition}\label{prp_four}
 In the four-party system ($n=4$), the unique symmetric pure-strategy Nash equilibrium is
\[
x^*
=
\left(
\frac{1-d}{2},\frac{1-d}{2},
\frac{1+d}{2},\frac{1+d}{2}
\right).
\]
\end{proposition}

The equilibrium differs from the Downsian benchmark. In the usual vote-maximization model, paired symmetric locations are at the first and third quartiles \citep{cox1987electoral}.\footnote{\cite{cox1987electoral} defines admissible objectives as a class of candidate objectives that includes not only vote-share maximization but also plurality-margin maximization and complete plurality maximization. He shows that his equilibrium characterization is robust as long as each candidate's objective belongs to this class. Maximization of the Myerson value does not belong to this class, because a party's payoff is determined not only by the distribution of vote shares but also by the policy profile, which induces the communication graph among parties. As a result, the equilibrium prediction can differ from that obtained under admissible objectives.} Here, parties maximize the Myerson value. When \(d<1/2\), locating at \((1/4,1/4,3/4,3/4)\) is not an equilibrium.\footnote{It is easily verified that \((1/4,1/4,3/4,3/4)\) constitutes an equilibrium if $d\geq 1/2$.} Instead, in  the equilibrium, the distance between the two party pairs is exactly equal to the communication threshold.

The intuition behind the proposition is easiest to see by starting from the
usual Downsian benchmark. In the present model,
however, this profile is not stable. At this profile, the two left parties form
one connected component and the two right parties form another connected component, but
the two components are disconnected from each other. Since each component controls exactly
one half of the seats, each component has value only \(q(1/2)=1/2\). Hence, each party
obtains Myerson value \(1/4\). A party can then profitably move slightly toward the center.
By doing so, it keeps the link with its partner, remains disconnected from the opposite pair,
and makes its own component strictly majoritarian. The component then has value \(1\), and
the deviating party obtains one half of this value. Thus, the usual Downsian benchmark
cannot be an equilibrium when coalition communication is restrictive. The same logic applies when the two party pairs are too far apart.

 By contrast, if the two pairs are too close, the
communication graph is complete (i.e., all four parties are connected). Thus, a party can move outward while remaining connected
to the other parties. Such a deviation increases its vote share and its pivotality in the
graph-restricted voting game. Therefore, neither excessive separation nor excessive
convergence can be stable.

The balance occurs when the distance between the two party pairs is exactly equal
to the coalition-communication threshold \(d\). At this point, moving inward reduces a
party's vote share without improving the value of the component the party belongs to, while moving outward breaks the
cross-pair communication links that are necessary to preserve its bargaining power. Hence, the two party pairs locate at the derived equilibrium.

As \(d\) becomes smaller, parties move closer to the median. On the one hand, as \(d\to 1/2\), the equilibrium converges to  \((1/4,1/4,3/4,3/4)\). On the other hand, as $d\to 0$, the equilibrium platforms converge to the median voter's ideal point; that is, the median voter theorem holds despite the multiparty setting. Although the details differ, these comparative statics are qualitatively the same as in the case of the three-party system.

At first glance, a smaller $d$ may seem likely to induce greater policy divergence because
policy compromise becomes more difficult. The analysis of the three- and four-party systems
shows the opposite: more restrictive coalition communication generates a centripetal force
that is absent from the standard vote-maximization benchmark.

\subsection{Application: Four-Party System with Two Extreme Parties}
To illustrate the applicability of the model, I conclude this section by considering an extension with a four-party system in which two parties are exogenously fixed at the two endpoints. This captures the idea that some ideologically extreme parties may not seek to maximize either vote shares or legislative bargaining power; instead, their primary objective may be to advocate their ideology. I then analyze how the remaining two mainstream parties, which maximize the Myerson value, choose their policy platforms in this environment.

Let \(A\) and \(B\) denote the fixed extreme parties, with $x_A=0$ and $x_B=1$. The remaining two mainstream parties, denoted by \(L\) and \(R\), choose their policy positions. I focus on symmetric profiles of the form $x=(0,s,1-s,1)$. 

In this setting, I obtain the following proposition. 
\begin{proposition}\label{prp_4_ext}
In the four-party system with two extreme parties, the set of symmetric pure-strategy Nash equilibria is
\[
X^*=\left\{(0,s,1-s,1)\ \middle|\ s>d,\ 1-2s\leq d,\ s<\frac12\right\}.
\]
\end{proposition}
Thus, the equilibrium is not unique in contrast to the previous subsection.
The two inequalities have distinct roles. The condition $1-2s\leq d$ links the two mainstream parties; the condition $s>d$ disconnects each mainstream party from the adjacent extreme party. Hence, in every equilibrium, $L$ and $R$ form the only nontrivial connected component. Their combined vote share is $1-s>1/2$, so each obtains Myerson value $1/2$. The mainstream bloc is therefore an equilibrium outcome rather than an exogenously imposed coalition structure, consistent with the logic of a \textit{cordon sanitaire} \citep{axelsen2024cordon,backlund2023government}.

The comparative statics are non-monotone over the full range of $d$. A lower $d$ makes it easier for a mainstream party to remain disconnected from the adjacent extreme, relaxing $s>d$, but it also requires the two mainstream parties to move closer together, tightening $1-2s\leq d$. The supremum of the distance between a mainstream platform and the median is
\[
\min\left\{\frac d2,\frac12-d\right\},
\]
which is largest at $d=1/3$ and converges to zero as either $d\to0$ or $d\to1/2$. Thus, the median voter theorem holds in both limiting cases: $d\to 0$ and $d\to 1/2$.

This prediction differs from vote-share maximization. With the extreme parties fixed at 0 and 1, every symmetric profile $(0,s,1-s,1)$ with $s\in(0,1/2]$ is a vote-share-maximizing equilibrium: a mainstream party's market share is locally independent of its interior location. Maximization of the Myerson value instead imposes the two coalition-connectivity restrictions above. 

Finally, from a different perspective, comparison with Proposition \ref{prp_two} highlights the role of extreme parties. Their presence can induce mainstream parties to diverge from the median voter’s ideal policy. However, this effect of extremism diminishes as $d$ is close to zero or close to a half.

\section{Concluding Remarks}
This paper embeds the Myerson value in a Downsian location game. A party's platform determines both its electoral weight and its position in a communication network, and these two channels jointly determine legislative bargaining power. The analysis yields three distinct predictions. In the two-party system, the usual Downsian centripetal logic obtains: the unique symmetric equilibrium is full convergence to the median. In the three-party system, there is a continuum of symmetric equilibria, and the communication constraint requires only that the two extreme parties be linked.  In the four-party case, the unique symmetric equilibrium places two parties at \((1-d)/2\) and two parties at \((1+d)/2\). Thus, more restrictive coalition communication (i.e., smaller $d$) generates a centripetal force in the three- and four-party systems. As an application, I also examined the four-party system with two extreme parties.

Overall, the results highlight the importance of explicitly incorporating coalition-power incentives into models of policy competition in multiparty parliamentary systems. They also demonstrate the usefulness of the Myerson value as a tractable way to capture such incentives.

Several extensions remain for future work. First, it is important to extend the analysis to general \(n\)-party systems. However, for \(n\geq 5\), the communication graph can contain multiple intermediate parties that serve as mediators between distant parties, which makes the analysis substantially more complex. Second, while the Myerson value allows communication through any indirect path between parties, it may be more realistic to impose an upper bound on the length of admissible communication paths \citep{kaneko2026k}. This is particularly relevant when general $n$-party systems are considered. Third, there are other power indices based on related ideas, such as the restricted Banzhaf index \citep{napel2012monotonicity}. Whether the present results extend to such indices remains an open question for future research.

\bibliography{bibtex}

\newpage
\appendix
\section{Omitted Proofs}

\subsection{Proof of Proposition \ref{prp_two}}
Consider a symmetric profile \(x(s)=(s,1-s)\), where \(0\leq s\leq 1/2\). If \(s=1/2\), both parties locate at the median and each party receives vote share \(1/2\). Each singleton coalition has value \(1/2\), and the grand coalition has value \(1\). Hence, each party's Myerson value is
\[
\frac{1}{2}\cdot\frac12+\frac{1}{2}\cdot\left(1-\frac12\right)=\frac12.
\]

Now, suppose one party deviates to \(y\neq 1/2\), while the other party remains at \(1/2\). If \(y<1/2\), the deviator's vote share is \((y+1/2)/2<1/2\). If \(y>1/2\), the deviator's vote share is \(1-(y+1/2)/2<1/2\). Thus, after any such deviation, the nondeviating party has strictly more than one half of the votes. The deviator has no positive marginal contribution: if it is first in the order, its singleton coalition is losing; if it is second in the order, the other party is already winning. Therefore, the deviator's Myerson value is zero. Hence, no deviation from \((1/2,1/2)\) is profitable.

It remains to show that no symmetric profile with \(s<1/2\) is an equilibrium. At \((s,1-s)\), both parties have vote share \(1/2\). Hence,
\[
v^g(\{1\})=v^g(\{2\})=\frac12.
\]
Moreover, regardless of whether the two parties are linked, 
\[
v^g(\{1,2\})=1.
\]
Indeed, if the two parties are linked, then \(v^g(\{1,2\})=v(\{1,2\})=1\); if they are not linked, then
\[
v^g(\{1,2\})=v(\{1\})+v(\{2\})=\frac12+\frac12=1.
\]
Therefore, party \(1\)'s Myerson value is
\[
\mu_1
=
\frac12\cdot \frac12
+
\frac12\cdot\left(1-\frac12\right)
=
\frac12.
\]
By symmetry, party \(2\)'s Myerson value is also \(1/2\).

Let the left party deviate to \(y=s+\varepsilon\), where \(0<\varepsilon<1-2s\). Then, the deviator remains to the left of the other party, and its vote share is
\[
\frac{y+(1-s)}{2}
=
\frac12+\frac{\varepsilon}{2}
>
\frac12.
\]
Thus, the deviator is a singleton winning party.  Hence, its Myerson value is \(1\), which is strictly larger than \(1/2\). Therefore, no symmetric profile with \(s<1/2\) is a Nash equilibrium. The unique symmetric Nash equilibrium is \((1/2,1/2)\).
\hspace*{\fill}$\Box$

\subsection{Proof of Proposition \ref{prp_three}}
Denote the three parties by \(L,M,R\), located at \(s,1/2,1-s\), respectively.

\paragraph{Case 1: \(1-2s\leq d\).} The communication graph is complete. If \(s<1/2\), the vote shares are
\[
m_L=m_R=\frac{s+1/2}{2}=\frac{1+2s}{4},
\qquad
m_M=\frac12-s.
\]
Since \(d<1\), the condition \(1-2s\leq d\) implies \(s>0\). Hence, each singleton coalition has less than one half of the votes, while every two-party coalition has strictly more than one half. If \(s=1/2\), all parties locate at the median and each party receives vote share \(1/3\), so again no singleton coalition is winning and every two-party coalition is winning. Therefore, in the complete graph, each party's marginal contribution is positive if and only if exactly one other party precedes it. Since the two one-player predecessor sets each have Shapley coefficient \(1/6\), each party obtains
\[
\frac{1}{6}+\frac{1}{6}=\frac13.
\]

We now show that no unilateral deviation is profitable. Consider any party that deviates to some \(y\in[0,1]\), and denote the deviator by \(D\). Let \(j\) and \(k\) be the two nondeviating parties. Since \(1-2s\leq d\), the two nondeviating parties are directly linked.

Next, the deviator cannot obtain a weak majority of votes by a unilateral deviation. To see this, first suppose a side party deviates, say party \(L\). The nondeviating parties remain at \(1/2\) and \(1-s\). If \(y<1/2\), then \(m_D=(y+1/2)/2<1/2\). If \(y=1/2\), the deviator shares the same position as party \(M\), and its vote share is less than \(1/2\). If \(1/2<y<1-s\), then
\[
m_D=\frac{(y+1-s)-(1/2+y)}{2}=\frac{1/2-s}{2}<\frac12.
\]
If \(y\geq 1-s\), the deviator is at or to the right of party \(R\), and its vote share is again less than \(1/2\). Thus, a deviating side party never obtains a weak majority. Next, suppose the middle party deviates. The nondeviating parties remain at \(s\) and \(1-s\). If \(y<s\), then \(m_D=(y+s)/2<1/2\). If \(s<y<1-s\), then
\[
m_D=\frac{(y+1-s)-(s+y)}{2}=\frac{1-2s}{2}<\frac12,
\]
where the strict inequality follows from \(s>0\). If \(y>1-s\), the deviator is the rightmost party and its vote share is less than \(1/2\). If \(y=s\), then the deviator shares the voters who are closest to \(s\) with the party already located at \(s\). Hence, its vote share is
\[
m_D=\frac12\cdot \frac{s+1/2}{2}
=\frac{1+2s}{8}<\frac12.
\]
Similarly, if \(y=1-s\), then the deviator shares the voters who are closest to \(1-s\) with the party already located there, and hence
\[
m_D=\frac12\cdot \frac{1/2+1-s}{2}
=\frac{3-2s}{8}<\frac12.
\]

Since the two nondeviating parties are linked and have aggregate vote share \(1-m_D>1/2\), the coalition \(\{j,k\}\) already has value \(1\) in the graph-restricted game. Therefore, the deviator's marginal contribution is zero for the empty predecessor set and also zero for the predecessor set \(\{j,k\}\). The only predecessor sets that can yield a positive marginal contribution are \(\{j\}\) and \(\{k\}\). Each of these predecessor sets has Shapley coefficient \(1/6\), and each marginal contribution is at most \(1\). Thus, any deviation yields
\[
\mu_D\leq \frac{1}{6}+\frac{1}{6}=\frac13.
\]
Since every party obtains \(1/3\) at the original profile, no unilateral deviation is profitable. Thus, any symmetric profile satisfying \(1-2s\leq d\) is a Nash equilibrium.

\paragraph{Case 2: \(1-2s>d\). } Parties \(L\) and \(R\) are not directly linked. We show that party \(L\) has a profitable deviation. There are two cases.

First, suppose \(1/2-s>d\). Then, party \(L\) is isolated from both other parties. Since \(m_L<1/2\), party \(L\)'s Myerson value is zero. Let \(L\) deviate to \(y=1/2\). After the deviation, the deviator and party \(M\) are located at the same position and are directly linked. Their connected component has aggregate vote share strictly greater than \(1/2\), while neither party is individually winning. The party \(R\) is not linked to this component. Therefore, the component \(\{D,M\}\) is a two-player unanimity component of value \(1\), and the deviator obtains \(1/2>0\). This deviation is profitable.

Second, suppose \(1/2-s\leq d\). Then, the communication graph is the path \(L-M-R\). We first bound \(L\)'s Myerson value. The only predecessor set that can give \(L\) a positive marginal contribution is \(\{M\}\). Indeed, adding \(L\) to \(\{R\}\) leaves \(L\) and \(R\) disconnected, while adding \(L\) to \(\{M,R\}\) does not change the value because \(\{M,R\}\) is already a winning connected coalition. Hence,
\[
\mu_L\leq \frac{1}{6}.
\]
Now, let \(L\) deviate to \(y=1/2\). Since \(1/2-s\leq d\), after this deviation the graph is complete. No singleton coalition is winning, while every two-party coalition is winning. Thus, the deviator's Myerson value is \(1/3\), which is strictly larger than \(1/6\). This deviation is profitable.

In both cases, a symmetric profile with \(1-2s>d\) cannot be a Nash equilibrium. 
\hspace*{\fill}$\Box$

\subsection{Proof of Proposition \ref{prp_four}}

\paragraph{Step 1:} We first show that no such profile with \(s<t\) can be a Nash equilibrium.

Suppose \(s<t\). Denote the four parties, from left to right, by $
L_1,\ L_2,\ R_1,\ R_2.$ 
The vote shares of \(L_1\) and \(R_2\) are
\[
A:=\frac{s+t}{2},
\]
and the vote shares of \(L_2\) and \(R_1\) are
\[
B:=\frac{1-s-t}{2}.
\]

Let $
\delta:=t-s$ and 
$\gamma:=1-2t.$
Then
\[
\begin{aligned}
|x_{L_1}-x_{L_2}|&=|x_{R_1}-x_{R_2}|=\delta,
&|x_{L_2}-x_{R_1}|&=\gamma,\\
|x_{L_1}-x_{R_1}|&=|x_{L_2}-x_{R_2}|=\delta+\gamma,
&|x_{L_1}-x_{R_2}|&=2\delta+\gamma.
\end{aligned}
\]

We now consider all possible communication graphs that can arise from such a symmetric
profile with \(s<t\).

\paragraph{Case 1: \(\gamma>d\).}
Parties \(L_2\) and \(R_1\) are not linked. If also \(\delta>d\), then all parties
are isolated. If \(\delta\leq d\), then the graph consists of two disconnected pairs,
\[
\{L_1,L_2\}
\quad\text{and}\quad
\{R_1,R_2\}.
\]
In either case, party \(L_1\)'s Myerson value is at most \(1/4\).

Let \(L_1\) deviate to $
y=t+\varepsilon,$
where
$0<\varepsilon<\min\{d,\gamma-d\}.$
Then the deviator is linked to \(L_2\), but not to \(R_1\). The connected component
containing the deviator and \(L_2\) has aggregate vote share
\[
\frac{y+(1-t)}{2}
=
\frac12+\frac{\varepsilon}{2}
>
\frac12.
\]
Hence, this component has value \(1\). Since neither party in this two-party component is
individually winning, the component is a two-player unanimity game of value \(1\). The
deviator therefore obtains
\[
\frac12>\frac14.
\]
Thus, no equilibrium can satisfy \(\gamma>d\).

\paragraph{Case 2: \(\gamma\leq d\) and \(\delta>d\).}
In this case, \(L_2\) and \(R_1\) are linked, but \(L_1\) and \(R_2\) are isolated from them.
Hence, the graph consists of
\[
\{L_2,R_1\},\qquad \{L_1\},\qquad \{R_2\}.
\]
Parties \(L_1\) and \(R_2\) have Myerson value zero.

First, suppose \(\gamma>0\). Since \(\delta>d\), choose \(\varepsilon>0\) such that
$
0<\varepsilon<\min\{\gamma,\delta-d\}.$
Let \(L_1\) deviate to
$
y=t-d+\varepsilon.$
Then the deviator is linked to \(L_2\), because \(t-y=d-\varepsilon<d\), but not to \(R_1\),
because
\[
(1-t)-y
=
\gamma+d-\varepsilon
>
d.
\]
The component containing the deviator and \(L_2\) has aggregate vote share exactly \(1/2\).
Therefore this component has value \(q(1/2)=1/2\), and the deviator obtains one half of this
component value:
\[
\frac14>0.
\]

If \(\gamma=0\), let \(L_1\) deviate to
\[
y=t-\frac d2.
\]
Then, the deviator is linked to both \(L_2\) and \(R_1\). Denote the deviator by \(D\). The
coalition \(\{L_2,R_1\}\) has aggregate vote share less than \(1/2\), whereas the connected
coalition \(\{D,L_2,R_1\}\) has aggregate vote share greater than \(1/2\). Thus \(D\) has a
positive marginal contribution for the predecessor set \(\{L_2,R_1\}\), and so its Myerson
value is strictly positive. Since \(L_1\)'s original Myerson value is zero, this deviation is
profitable.

Thus, no equilibrium can satisfy \(\gamma\leq d\) and \(\delta>d\).

\paragraph{Case 3: \(\gamma\leq d\), \(\delta\leq d\), and \(\delta+\gamma>d\).}
In this case, the communication graph is the path
\[
L_1-L_2-R_1-R_2.
\]
We first compute \(L_1\)'s Myerson value. The Shapley coefficient is \(1/12\) for each
one-player or two-player predecessor set.

If \(A<B\), the only predecessor sets for which \(L_1\) has a positive marginal contribution are
\[
\{L_2\}
\quad\text{and}\quad
\{L_2,R_2\}.
\]
For each of these predecessor sets, adding \(L_1\) creates the connected component
\(\{L_1,L_2\}\), whose aggregate vote share is \(A+B=1/2\). Hence the marginal contribution
is \(1/2\) in each case, and
\[
\mu_{L_1}
=
\frac{1}{12}\cdot\frac12
+
\frac{1}{12}\cdot\frac12
=
\frac{1}{12}.
\]

If \(A=B\), there is one additional positive marginal contribution, namely for the predecessor
set
\[
\{L_2,R_1\}.
\]
In that case, \(\{L_2,R_1\}\) has aggregate vote share \(1/2\), whereas
\(\{L_1,L_2,R_1\}\) has aggregate vote share \(3/4\). Thus the marginal contribution is
\(1/2\). Therefore,
\[
\mu_{L_1}
=
3\cdot \frac{1}{12}\cdot\frac12
=
\frac18.
\]

If \(A>B\), the positive marginal contributions arise for
\[
\{L_2\},\quad
\{L_2,R_1\},\quad
\{L_2,R_2\}.
\]
The marginal contributions are, respectively,
\[
\frac12,\quad 1,\quad \frac12.
\]
Therefore,
\[
\mu_{L_1}
=
\frac{1}{12}
\left(
\frac12+1+\frac12
\right)
=
\frac16.
\]
By symmetry, the same calculation applies to \(R_2\). Hence
\[
\mu_{L_1}=\mu_{R_2}
=
\begin{cases}
1/12 & \text{if } A<B,\\[1mm]
1/8  & \text{if } A=B,\\[1mm]
1/6  & \text{if } A>B.
\end{cases}
\]
In all cases,
\[
\mu_{L_1}<\frac14.
\]

We now show that \(L_1\) has a profitable deviation. Let \(L_1\) deviate to \(y=t\). Denote the deviator by \(D\), the other party at \(t\) by \(L\), and the parties at \(1-t\) and \(1-s\) by \(R_1\) and \(R_2\), respectively. After the deviation, the vote shares are
\[
m_D=m_L=\frac14,
\qquad
m_{R_1}=B,
\qquad
m_{R_2}=A.
\]
Since \(\gamma\leq d\), the parties \(D\), \(L\), and \(R_1\) are pairwise connected. Since \(\delta\leq d\), \(R_1\) is linked to \(R_2\). Finally, since \(\delta+\gamma>d\), neither \(D\) nor \(L\) is directly linked to \(R_2\).

The positive marginal contributions of \(D\) are as follows. For the predecessor sets \(\{L\}\), \(\{L,R_2\}\), and \(\{R_1,R_2\}\), the marginal contribution is \(1/2\). For the predecessor set \(\{R_1\}\), the marginal contribution is \(1\), \(1/2\), or \(0\) depending on \(A<B\), \(A=B\), or \(A>B\). For the predecessor set \(\{L,R_1\}\), the marginal contribution is \(0\), \(1/2\), or \(1\) according as \(A<B\), \(A=B\), or \(A>B\). Hence the marginal contributions for \(\{R_1\}\) and \(\{L,R_1\}\) always sum to \(1\). All other marginal contributions are zero. Therefore,
\[
\mu_D
=
\frac{1}{12}
\left(
\frac12+\frac12+\frac12+1
\right)
=
\frac{5}{24}
>
\frac16
\geq
\mu_{L_1}.
\]
Thus, no equilibrium can satisfy \(\gamma\leq d\), \(\delta\leq d\), and \(\delta+\gamma>d\).

\paragraph{Case 4: \(\delta+\gamma\leq d\) and \(2\delta+\gamma>d\).}
In this case, all links are present except the link between \(L_1\) and \(R_2\). Since
\[
\delta+\gamma=1-s-t=2B\leq d<\frac12,
\]
we have \(B<1/4\), and hence \(A>1/4\).

We first compute the Myerson values at the original profile. Consider \(L_1\). The positive
marginal contributions of \(L_1\) arise for the following predecessor sets:
\[
\{L_2\},\quad
\{R_1\},\quad
\{L_2,R_1\},\quad
\{L_2,R_2\},\quad
\{R_1,R_2\}.
\]
The corresponding marginal contributions are
\[
\frac12,\quad
\frac12,\quad
1,\quad
\frac12,\quad
\frac12.
\]
Indeed, adding \(L_1\) to \(\{L_2\}\) or \(\{R_1\}\) creates a connected coalition with aggregate
vote share \(A+B=1/2\). Adding \(L_1\) to \(\{L_2,R_1\}\) creates a connected coalition with
aggregate vote share \(A+2B>1/2\), while \(\{L_2,R_1\}\) itself has aggregate vote share
\(2B<1/2\). Finally, adding \(L_1\) to either \(\{L_2,R_2\}\) or \(\{R_1,R_2\}\) changes the
value of the connected component from \(1/2\) to \(1\). Hence,
\[
\mu_{L_1}
=
\frac{1}{12}
\left(
\frac12+\frac12+1+\frac12+\frac12
\right)
=
\frac14.
\]
By symmetry, \(\mu_{R_2}=1/4\).

Similarly, for \(L_2\), the positive marginal contributions arise for
\[
\{L_1\},\quad
\{R_2\},\quad
\{L_1,R_1\},\quad
\{L_1,R_2\},\quad
\{R_1,R_2\}.
\]
The corresponding marginal contributions are
\[
\frac12,\quad
\frac12,\quad
\frac12,\quad
1,\quad
\frac12.
\]
Thus,
\[
\mu_{L_2}
=
\frac{1}{12}
\left(
\frac12+\frac12+\frac12+1+\frac12
\right)
=
\frac14.
\]
By symmetry, \(\mu_{R_1}=1/4\). Therefore,
\[
\mu_i=\frac14
\qquad
\text{for every party } i.
\]

Now, let \(L_1\) deviate to
$
y=s+\varepsilon,
$
where
$
0<\varepsilon<\min\{\delta,\,2\delta+\gamma-d,\,2B\}.$
The first inequality ensures that the deviator remains to the left of \(L_2\). The second
inequality ensures that the deviator is still not linked to \(R_2\), because
\[
(1-s)-y
=
2\delta+\gamma-\varepsilon
>
d.
\]
Thus, the communication graph remains the complete graph minus the edge between the
deviator and \(R_2\).

Denote the deviator by \(D\). After the deviation, the vote shares are
\[
m_D=A+\frac{\varepsilon}{2},
\qquad
m_{L_2}=B-\frac{\varepsilon}{2},
\qquad
m_{R_1}=B,
\qquad
m_{R_2}=A.
\]
The positive marginal contributions of \(D\) arise for the predecessor sets
\[
\{L_2\},\quad
\{R_1\},\quad
\{L_2,R_1\},\quad
\{L_2,R_2\},\quad
\{R_1,R_2\}.
\]
The corresponding marginal contributions are
\[
\frac12,\quad
1,\quad
1,\quad
1,\quad
\frac12.
\]
To see this, note that \(D\) together with \(L_2\) has aggregate vote share \(1/2\), whereas
\(D\) together with \(R_1\) has aggregate vote share strictly greater than \(1/2\). Moreover,
adding \(D\) to \(\{L_2,R_1\}\) or \(\{L_2,R_2\}\) makes the connected component containing
\(D\) strictly majoritarian. Finally, \(\{R_1,R_2\}\) has aggregate vote share \(1/2\), and adding
\(D\) changes the value from \(1/2\) to \(1\). Hence
\[
\mu_D
=
\frac{1}{12}
\left(
\frac12+1+1+1+\frac12
\right)
=
\frac13
>
\frac14.
\]
Therefore, the original profile is not an equilibrium.

\paragraph{Case 5: \(2\delta+\gamma\leq d\).}
In this case, the communication graph is complete. Because
\[
2\delta+\gamma=1-2s\leq d<\frac12,
\]
we have \(s>1/4\), and hence
\[
A=\frac{s+t}{2}>\frac14>B.
\]
Thus, \(L_1\) and \(R_2\) have larger vote shares than \(L_2\) and \(R_1\). Since the graph is
complete, the Myerson value coincides with the Shapley--Shubik index of the weighted
voting game with weights
\[
(A,B,B,A),
\qquad A+B=\frac12,
\qquad A>B.
\]

We first compute these values explicitly. Consider \(L_1\). Its positive marginal contributions
arise for
\[
\{L_2\},\quad
\{R_1\},\quad
\{R_2\},\quad
\{L_2,R_1\},\quad
\{L_2,R_2\},\quad
\{R_1,R_2\}.
\]
The corresponding marginal contributions are
\[
\frac12,\quad
\frac12,\quad
1,\quad
1,\quad
\frac12,\quad
\frac12.
\]
Therefore,
\[
\mu_{L_1}
=
\frac{1}{12}
\left(
\frac12+\frac12+1+1+\frac12+\frac12
\right)
=
\frac13.
\]
By symmetry, \(\mu_{R_2}=1/3\).

Next, consider \(L_2\). Its positive marginal contributions arise for
\[
\{L_1\},\quad
\{R_2\},\quad
\{L_1,R_1\},\quad
\{R_1,R_2\}.
\]
For each of these predecessor sets, the marginal contribution is \(1/2\). Hence
\[
\mu_{L_2}
=
4\cdot \frac{1}{12}\cdot\frac12
=
\frac16.
\]
By symmetry, \(\mu_{R_1}=1/6\). Thus
\[
\mu_{L_1}=\mu_{R_2}=\frac13,
\qquad
\mu_{L_2}=\mu_{R_1}=\frac16.
\]

We show that \(L_2\) has a profitable deviation. Let \(L_2\) deviate to
$
y=t-\varepsilon,$
where \(\varepsilon>0\) is sufficiently small that the communication graph remains complete and
\[
B+\frac{\varepsilon}{2}<\frac14
\quad\text{and}\quad
A-\frac{\varepsilon}{2}>\frac14.
\]
Such an \(\varepsilon\) exists because \(A>1/4>B\). After this deviation, the deviator's vote
share is \(B+\varepsilon/2\). Denote the deviator by \(D\). The vote shares are
\[
m_{L_1}=A-\frac{\varepsilon}{2},
\qquad
m_D=B+\frac{\varepsilon}{2},
\qquad
m_{R_1}=B,
\qquad
m_{R_2}=A.
\]
Because the graph remains complete, the Myerson value is again the Shapley--Shubik value.
The positive marginal contributions of \(D\) arise for
\[
\{L_1\},\quad
\{R_2\},\quad
\{L_1,R_1\},\quad
\{R_1,R_2\}.
\]
The corresponding marginal contributions are
\[
\frac12,\quad
1,\quad
1,\quad
\frac12.
\]
Therefore,
\[
\mu_D
=
\frac{1}{12}
\left(
\frac12+1+1+\frac12
\right)
=
\frac14
>
\frac16.
\]
Hence the original profile is not an equilibrium.

The five cases exhaust all possible communication graphs for a symmetric profile with
\(s<t\). Therefore, no symmetric profile with \(s<t\) can be a Nash equilibrium. Hence, any
symmetric pure-strategy equilibrium must satisfy
\(s=t\).

\paragraph{Step 2:} Thus, we focus on strategy profile:
$x(s)=(s,s,1-s,1-s).$  Each party receives vote share \(1/4\). Let \(\Delta:=1-2s\).

First, suppose \(\Delta>d\). Then, the communication graph has two components, each containing two parties. Thus, as in the example, each party's Myerson value is \(1/4\).

Now, let a party at the left position deviate from \(s\) to \(y=s+\varepsilon\), where
\[
0<\varepsilon<\min\{d,\Delta-d\}.
\]
The deviating party remains linked to the other left party, but it is still not linked to the two right parties. The left component now has aggregate vote share
\[
\frac{y+(1-s)}{2}
=
\frac12+\frac{\varepsilon}{2}
>
\frac12.
\]
Hence, the left component has value \(1\). The two left parties form a two-player unanimity component of value \(1\), so the deviating party obtains
\[
\frac12>
\frac14.
\]
Thus no equilibrium can satisfy \(\Delta>d\).

Next, suppose \(\Delta<d\). The communication graph is complete. Again each party initially obtains \(1/4\). Let a party at the left position deviate to \(y=s-\varepsilon\), where \(\varepsilon>0\) is small enough that \(\Delta+\varepsilon<d\). The graph remains complete. Denote the deviating party by \(D\), the other left party by \(L\), and the two right parties by \(R_1,R_2\). The vote shares after the deviation satisfy
\[
m_D=\frac{y+s}{2}>\frac14,
\qquad
m_D+m_L=\frac12,
\qquad
m_{R_1}=m_{R_2}=\frac14.
\]
We compute \(D\)'s Myerson value, which is the Shapley value because the graph is complete. The marginal contribution of \(D\) is positive only for the following predecessor sets:
\[
\{L\},\quad
\{R_1\},\quad
\{R_2\},\quad
\{L,R_1\},\quad
\{L,R_2\},\quad
\{R_1,R_2\}.
\]
The marginal contributions are, respectively,
\[
\frac12,
\quad
1,
\quad
1,
\quad
1,
\quad
1,
\quad
\frac12.
\]
The Shapley coefficient for any one-player predecessor set is \(1/12\), and the coefficient for any two-player predecessor set is also \(1/12\). Therefore
\[
\mu_D
=
\frac{1}{12}\left(\frac12+1+1+1+1+\frac12\right)
=
\frac{5}{12}
>
\frac14.
\]
Thus, no equilibrium can satisfy \(\Delta<d\).

It remains to consider the case \(\Delta=d\). Let
\[
a:=\frac{1-d}{2},
\qquad
b:=\frac{1+d}{2}.
\]
The candidate profile is \(x^*=(a,a,b,b)\). The communication graph is complete. Each party has vote share \(1/4\). Thus, a party's marginal contribution is \(0\) if it is first in the order, \(1/2\) if it is second, \(1/2\) if it is third, and \(0\) if it is fourth. Hence, each party obtains
\[
\frac14\left(0+\frac12+\frac12+0\right)
=
\frac14.
\]

We now show that no unilateral deviation is profitable. By symmetry, consider a party initially located at \(a\), and let it deviate to \(y\). The other three parties remain at \(a,b,b\).

If \(y<a-d\), the deviating party is isolated. Its vote share is less than \(1/2\), so its value is \(0\).

If \(a-d\leq y<a\), the deviating party \(D\) is directly linked to the remaining party \(L\) at \(a\), while \(L\) is directly linked to the two parties \(R_1,R_2\) at \(b\). Thus, although \(D\) is not directly linked to \(R_1\) or \(R_2\), it can be connected to them through \(L\) when \(L\) is in the predecessor set. Let \(w:=m_D=(y+a)/2\). Then \(m_L=1/2-w\) and \(m_{R_1}=m_{R_2}=1/4\). The marginal contribution of \(D\) is \(1/2\) for the predecessor set \(\{L\}\). For each of the predecessor sets \(\{L,R_1\}\) and \(\{L,R_2\}\), the marginal contribution is at most \(1\). For all other predecessor sets, the marginal contribution is zero. Therefore,
\[
\mu_D
\leq
\frac{1}{12}\cdot\frac12
+2\cdot\frac{1}{12}
=
\frac{5}{24}
<
\frac14.
\]

If \(a<y<b\), the communication graph is complete. Let \(D\) be the deviator, \(L\) the remaining party at \(a\), and \(R_1,R_2\) the two parties at \(b\). The vote shares are
\[
m_D=\frac d2,
\qquad
m_L=\frac{a+y}{2},
\qquad
m_{R_1}=m_{R_2}=\frac{2-y-b}{4}.
\]
The predecessor set \(\{L\}\) gives marginal contribution \(1\), because \(m_L<1/2\) and \(m_L+m_D=(y+b)/2>1/2\). The predecessor set \(\{R_1,R_2\}\) also gives marginal contribution \(1\), because \(m_{R_1}+m_{R_2}<1/2\) and \(m_{R_1}+m_{R_2}+m_D=(2-y-a)/2>1/2\), where the last inequality follows from \(y<b=1-a\). By contrast, neither \(\{R_1\}\) nor \(\{R_2\}\) gives a positive marginal contribution. Indeed, for \(i=1,2\),
\[
m_{R_i}+m_D<\frac12
\quad\Longleftrightarrow\quad
 y>b-2a,
\]
and this inequality holds because \(b-2a<a<y\) when \(d<1/2\). The predecessor sets \(\{L,R_1\}\) and \(\{L,R_2\}\) also give zero marginal contribution, because they already have value \(1\) before \(D\) is added. To see this, for \(i=1,2\),
\[
m_L+m_{R_i}>\frac12
\quad\Longleftrightarrow\quad
 y>b-2a,
\]
which again follows from \(b-2a<a<y\). Therefore, the only predecessor sets for which \(D\) has a positive marginal contribution are
\[
\{L\}
\quad\text{and}\quad
\{R_1,R_2\}.
\]
Since each has Shapley coefficient \(1/12\), we obtain
\[
\mu_D=\frac{1}{12}+\frac{1}{12}=\frac16<\frac14.
\]

If \(y=b\), then the remaining party at \(a\) has vote share \(1/2\), and the three parties at \(b\) each have vote share \(1/6\). The graph is complete. For the deviator \(D\), the predecessor set \(\{L\}\), where \(L\) is the party at \(a\), gives marginal contribution \(1/2\), because adding \(D\) changes the coalition value from \(1/2\) to \(1\). The predecessor set consisting of the other two parties at \(b\) also gives marginal contribution \(1/2\), because adding \(D\) changes the coalition value from \(0\) to \(1/2\). All other predecessor sets give zero marginal contribution. Therefore
\[
\mu_D=
\frac{1}{12}\cdot\frac12
+
\frac{1}{12}\cdot\frac12
=
\frac{1}{12}<\frac14.
\]

If \(b<y\leq b+d\), the deviator is linked to the two parties at \(b\), but not directly to the party at \(a\). The only predecessor set that gives a positive marginal contribution is the set consisting of the two parties at \(b\). When they precede the deviator, adding the deviator makes the right component have aggregate vote share exactly \(1/2\), so the marginal contribution is \(1/2\). Hence
\[
\mu_D=
\frac{1}{12}\cdot\frac12
=
\frac{1}{24}<\frac14.
\]
Finally, if \(y>b+d\), the deviator is isolated and has Myerson value \(0\). Thus, no deviation yields a payoff above \(1/4\).  \(x^*\) is therefore a Nash equilibrium. \hspace*{\fill}$\Box$

\subsection{Proof of Proposition \ref{prp_4_ext}}

When \(0<s<1/2\), the vote shares are
\[
m_A=m_B=\frac{s}{2},
\qquad
m_L=m_R=\frac{1-s}{2}.
\]

\paragraph{Step 1.}
First, suppose that
\[
s>d,\qquad 1-2s\leq d,\qquad s<\frac12.
\]
Then, \(L\) and \(R\) are directly linked. On the other hand, \(A\) is not linked to \(L\), and \(R\) is not linked to \(B\). Thus, the only nontrivial connected component is \(\{L,R\}\). The aggregate vote share of \(\{L,R\}\) is
\[
m_L+m_R=1-s>\frac12.
\]
Hence, this component has value \(1\). Since neither \(L\) nor \(R\) is individually winning, 
\[
\mu_L=\mu_R=\frac12.
\]

We now show that no unilateral deviation is profitable. By symmetry, consider a deviation by \(L\). Let \(D\) denote the deviating party and let \(y\in[0,1]\) be its new position.

First, suppose \(y<1-s\). The deviator \(D\) cannot be linked to \(B\), because
$1-y>1-(1-s)=s>d.$
Thus, \(D\) can be linked only to \(A\), to \(R\), to both, or to neither.

If \(D\) is linked to neither \(A\) nor \(R\), then \(D\) is isolated. Since its vote share is less than \(1/2\), its Myerson value is \(0\). If \(D\) is linked only to \(A\), then \(y\leq d<s\), and hence
\[
m_A+m_D=\frac{y+1-s}{2}<\frac12.
\]
Thus, the component containing \(D\) is losing, and again \(\mu_D=0\). If \(D\) is linked only to \(R\), then the component \(\{D,R\}\) has aggregate vote share greater than \(1/2\), while neither \(D\) nor \(R\) is individually winning. Hence \(\{D,R\}\) is a two-player unanimity component of value \(1\), and
\[
\mu_D=\frac12.
\]
Finally, suppose that \(D\) is linked to both \(A\) and \(R\). Then, the connected component containing \(D\) is $\{A, D, R\}$.
Because \(y\leq d<s\), the component \(\{A,D\}\) is losing. By contrast, the component \(\{D,R\}\) is winning. Hence, \(D\) has a positive marginal contribution only when \(R\) is already in the predecessor set. That is, the positive marginal contributions arise for
\[
\{R\},\quad \{A,R\},\quad \{R,B\},\quad \{A,R,B\}.
\]
In each case, the marginal contribution is \(1\). Therefore,
\[
\mu_D
=
\frac{1}{12}
+
\frac{1}{12}
+
\frac{1}{12}
+
\frac14
=
\frac12.
\]
Thus, for any deviation with \(y<1-s\),
\[
\mu_D\leq \frac12.
\]

Next, suppose \(y>1-s\). In this case, \(D\) cannot be linked to \(A\), and it can be linked only to \(R\), to \(B\), to both, or to neither. If \(D\) is linked to neither \(R\) nor \(B\), or only to \(B\), then the component containing \(D\) is losing and \(\mu_D=0\). If \(D\) is linked only to \(R\),  the component $\{D, R\}$ wins, but $R$'s vote share is smaller than $1/2$ because $m_R=y/2.$ Thus, $\mu_D=1/2$. If \(D\) is linked to both \(R\) and \(B\), then \(D\)'s positive marginal contributions arise exactly when \(R\) is already in the predecessor set, and the same calculation gives $\mu_D=1/2$. Therefore, for any deviation with \(y>1-s\),
\[
\mu_D\leq \frac12.
\]

Finally, suppose \(y=1-s\). Then \(D\) and \(R\) are located at the same position. The total vote share of the location \(1-s\) is \(1/2\), and \(D\) and \(R\) split this vote share equally. Hence
\[
m_D=m_R=\frac14.
\]
The component \(\{D,R\}\) has value \(1/2\), and \(D\)'s Myerson value is
\[
\mu_D=\frac14<\frac12.
\]
Therefore, no unilateral deviation gives a payoff strictly greater than \(1/2\). Since \(L\) and \(R\) obtain \(1/2\) at the original profile, the profile is a Nash equilibrium.

\paragraph{Step 2.} It remains to show that no other symmetric profile is an equilibrium. First, suppose \(s=1/2\). Then, the component \(\{L,R\}\) has aggregate vote share \(1/2\). Hence, each of \(L\) and \(R\) obtains
\[
\mu_L=\mu_R=\frac14.
\]
Now, let \(L\) deviate to \(y=1/2-\varepsilon\), where \(\varepsilon>0\) is sufficiently small that \(D\) remains linked to \(R\) but is not linked to \(A\). Then \(\{D,R\}\) has aggregate vote share strictly greater than \(1/2\), while neither \(D\) nor \(R\) is individually winning. Thus
\[
\mu_D=\frac12>\frac14.
\]
Therefore, \(s=1/2\) is not an equilibrium.

Next, suppose \(0<s<1/2\) and at least one of the two conditions $
s>d$ and $1-2s\leq d$ fails. There are three cases.

First, suppose \(s\leq d\) and \(1-2s>d\). Then, the graph consists of two disconnected pairs,
\[
\{A,L\}
\quad\text{and}\quad
\{R,B\}.
\]
Each pair has aggregate vote share \(1/2\), and the Myerson value of \(L\) is
\[
\mu_L=\frac14.
\]

Second, suppose \(s\leq d\) and \(1-2s\leq d\). Then, all parties are connected. A direct calculation gives
\[
\mu_L=\frac{5}{12}<\frac12.
\]

Third, suppose \(s>d\) and \(1-2s>d\). Then, \(L\) is not linked to any other party. Since \(L\) is not individually winning,
\[
\mu_L=0.
\]

In all three cases,
\[
\mu_L<\frac12.
\]
We now construct a profitable deviation. Let \(L\) deviate to a point \(y\) slightly to the left of \(1-s\), chosen so that
\[
1-s-d\leq y<1-s,
\qquad
d<y<1-d.
\]
Such a \(y\) exists. Then, the deviator \(D\) is linked to \(R\), but not to \(A\) or \(B\). The component \(\{D,R\}\) has aggregate vote share strictly greater than \(1/2\), while neither \(D\) nor \(R\) is individually winning. Furthermore, $m_R+m_B<1/2$ holds. Hence,
\[
\mu_D=\frac12,
\]
which is strictly larger than the original payoff of \(L\). Therefore, no such profile is an equilibrium.

Finally, suppose \(s=0\). Then, \(L\) is located at \(0\) and \(R\) is located at \(1\).  Thus, $\mu_L=1/4$. If \(L\) deviates to \(y=1/2\), then the deviator is isolated but has vote share \(1/2\), so its Myerson value is \(1/2\). This deviation is profitable. Hence \(s=0\) is not an equilibrium.

Combining these arguments, we obtain the proposition. \hspace*{\fill}$\Box$
\end{spacing}

\end{document}